\documentclass[12pt,a4paper]{article}

\usepackage[british]{babel}

\usepackage[a4paper,top=2cm,bottom=2cm,left=2.5cm,right=2.5cm,marginparwidth=1.75cm]{geometry}

\usepackage[style=nature, backend=biber]{biblatex} 
\DeclareLanguageMapping{british}{british-apa} 
\DeclareFieldFormat[article]{volume}{\textbf{#1}} 

\usepackage{amsmath}
\usepackage{graphicx}
\usepackage[colorlinks=true, allcolors=blue]{hyperref}
\usepackage{orcidlink}
\usepackage[title]{appendix}
\usepackage{mathrsfs}
\usepackage{amsfonts}
\usepackage{booktabs} 
\usepackage{caption}  
\usepackage{threeparttable} 
\usepackage{algorithm}
\usepackage{algorithmicx}
\usepackage{algpseudocode}
\usepackage{listings}
\usepackage{enumitem}
\usepackage{chngcntr}
\usepackage{booktabs}
\usepackage{lipsum}
\usepackage{subcaption}
\usepackage{authblk}
\usepackage[T1]{fontenc}    
\usepackage{csquotes}       
\usepackage{diagbox}
\usepackage{xcolor}
\usepackage{pdfpages}
\usepackage[normalem]{ulem}
\newcommand\redsout{\bgroup\markoverwith{\textcolor{red}{\rule[0.5ex]{2pt}{0.75pt}}}\ULon}

\usepackage{setspace}
\usepackage{titlesec}
\titleformat{\section} 
  {\normalfont\Large\bfseries}{\thesection.}{1em}{}
  
\usepackage{float}   
\usepackage{caption} 


\newcommand{\revnew}[1]{\textcolor{black}{#1}}
\title{Distributed Quantum Approximate Optimization Algorithm on a Quantum-Centric Supercomputing Architecture}

\author[1*]{Seongmin Kim}
\author[2]{Vincent R. Pascuzzi}
\author[1,3]{Zhihao Xu}
\author[3*]{Tengfei Luo}
\author[4*]{Eungkyu Lee}
\author[1*]{In-Saeng Suh}

\affil[1]{\small National Center for Computational Sciences, Oak Ridge National Laboratory, Oak Ridge, Tennessee 37830, United States}
\affil[2]{\small IBM Quantum, IBM T.J. Watson Research Center, Yorktown Heights, NY 10598}
\affil[3]{\small Department of Aerospace and Mechanical Engineering, University of Notre Dame, Notre Dame, Indiana 46556, United States}
\affil[4]{\small Department of Electronic Engineering, Kyung Hee University, Yongin-Si, Gyeonggi-do 17104, Republic of Korea}

\date{}  

\begin{document}
\maketitle
*Corresponding authors. Email: kims@ornl.gov, tluo@nd.edu, eleest@khu.ac.kr, and suhi@ornl.gov
\\
\begin{abstract}
Quantum approximate optimization algorithm (QAOA) has shown promise in solving combinatorial optimization problems by providing quantum speedup on near-term gate-based quantum computing systems. However, QAOA faces challenges for high-dimensional problems due to the large number of qubits required and the complexity of deep circuits, limiting its scalability for real-world applications. In this study, we present a distributed QAOA (DQAOA), which leverages distributed computing strategies to decompose a large computational workload into smaller tasks that require fewer qubits and shallower circuits than necessitated to solve the original problem. These sub-problems are processed using a combination of high-performance and quantum computing resources. The global solution is iteratively updated by aggregating sub-solutions, allowing convergence toward the optimal solution. We demonstrate that DQAOA can handle considerably large-scale optimization problems (e.g., 1,000-bit problem), achieving a high \revnew{solution quality} and short time-to-solution, outperforming existing strategies. Furthermore, we realize DQAOA on a quantum-centric supercomputing architecture, paving the way for practical applications of gate-based quantum computers in real-world optimization tasks. To extend DQAOA's applicability to materials science, we further develop an active learning algorithm integrated with our DQAOA (AL-DQAOA), which involves machine learning, DQAOA, and active data production in an iterative loop. We successfully optimize photonic structures using AL-DQAOA, indicating that solving real-world optimization problems using gate-based quantum computing is feasible. We expect the proposed DQAOA to be applicable to a wide range of optimization problems and AL-DQAOA to find broader applications in material design.
\end{abstract}


\section{Introduction}

Quantum computing, which leverages quantum phenomena such as superposition and entanglement, promises to solve complex problems that appear classically intractable \cite{graham2022multi, daley2022practical, huang2022quantum, arute2019quantum}. This transformative technology harnesses the principles of quantum mechanics to process information in fundamentally new ways, revolutionizing various fields including cryptography, materials science, and artificial intelligence \cite{bernstein2017post, ren2022experimental, west2023towards, furrutter2024quantum}. Researchers are increasingly utilizing quantum computing to accelerate their work across various domains \cite{bernal2022perspectives, akahoshi2024partially}. In particular, the quantum approximate optimization algorithm (QAOA) \cite{farhi2019qaoa}, a variational quantum algorithm that leverages both classical and quantum computing to optimize a cost function, has shown particular promise. QAOA iteratively adjusts parameters in both its classical and quantum components (Fig.~\ref{fig:fig-1}a, see details in the Methods section) and has demonstrated potential for efficiently addressing complex optimization problems~\cite{cerezo2021variational, lykov2023sampling}. For example, a recent study by Shaydulin et al. has verified that QAOA can outperform the best-known classical solver for specific optimization problems \cite{shaydulin2024evidence}. Furthermore, other researchers have also applied this quantum algorithm to various combinatorial optimization problems such as the traveling salesman problem, Max-Cut problem, and knapsack problem, demonstrating potential advantages of leveraging quantum computing \cite{zhou2023qaoa, van2021quantum, zhou2020quantum}.

To improve the performance of QAOA, various strategies are studied, such as advanced ansatz preparation, optimized initial parameters, and optimized number of layers \cite{brandhofer2022benchmarking, pellow2024effect}. However, there are several challenges when employing QAOA in practice on today's quantum devices. The biggest challenge is that current quantum computing systems are characterized by a limited number of qubits and can execute relatively low-depth circuits \cite{preskill2018quantum, de2021materials, bennewitz2022neural}. These limitations make it challenging to scale up to real-world problems (e.g., optimization of photonic structures), which generally require a large number of qubits to represent high-dimensional binary spaces with deep circuits. Consequently, most QAOA studies involve small problems requiring only a few to tens of qubits \cite{lykov2023fast, zhou2023qaoa, decross2023qubit}. For example, Kim et al. demonstrated that QAOA running with Qiskit Aer simulators exhibits a low \revnew{solution quality} and long computation times for problems requiring more than 28 qubits due to an unoptimized number of layers, unoptimized initial parameters, and the barren plateau problem, highlighting its difficulty in solving large-scale optimization problems \cite{kim2024performance}. 

The research community has proposed QUBO partition strategies to address such challenges. However, these approaches are primarily focused on sparse problems involving a small number of interactions such as Max-Cut, thus they are less effective when applied to real-world problems characterized by large and dense configurations \cite{dunjko2018computational, li2022large, zhou2023qaoa}. Another promising approach is to leverage hybrid architectures that integrate high-performance computing (HPC) with quantum computing (QC). Here, HPC systems distribute a large job into numerous sub-tasks through distributed computing across multi-cores/nodes, leaving the QC to execute circuits amenable to today's devices. In this way, a quantum-centric supercomputing (QCSC) platform can leverage more effectively current quantum systems by distributing heavy workloads and optimizing the use of limited quantum resources \cite{alexeev2024quantum}. Consequently, QCSC can enhance the overall computational capability by leveraging the strengths of both technologies \cite{britt2017high, humble2021quantum, lykov2023fast}. However, there has been a lack of research on integrating HPC and QC to solve large-scale optimization problems, especially those beyond the capabilities of current quantum computers. 

In this work, we develop a distributed QAOA (DQAOA) that leverages a QCSC-type architecture to efficiently solve large-scale real-world optimization problems (e.g., optimizing metamaterial structures). Our algorithm achieves a high \revnew{solution quality (i.e., low 1-approximation ratio (AR) value)} and exceptional time-to-solution for large-scale problems, outperforming conventional quantum algorithms, including the divide-and-conquer method. Additionally, we execute DQAOA across multiple quantum devices, demonstrating its feasibility for scalable today's quantum hardware applications. We further propose an active learning algorithm integrated with DQAOA for material design, successfully applying it to optimize high-performance photonic structures. This demonstrates the capability of our approach to address real-world, large-scale problems using current quantum devices. Furthermore, by combining the classical computational power of HPC systems with QC, our approach tackles key challenges in materials science, paving the way for practical and effective applications of QC in solving complex optimization tasks. 



\section{Results and Discussion}
\subsection{Distributed QAOA}
Combinatorial optimization problems, categorized as nondeterministic polynomial-time (\textit{NP})-hard problems, can be mapped to quadratic unconstrained binary optimization (QUBO) formulations. These QUBOs can serve as cost Hamiltonians for QAOA \cite{date2021qubo, kim2022high, kim2023design, kim2024wide}. However, translating large and dense QUBOs into QAOA circuits requires a substantial number of two-qubit gates to encode the pairwise interactions between variables, leading to deep circuit depths. Furthermore, the number of qubits needed scales with QUBO sizes, presenting challenges for larger problems. While QAOA has shown promise in solving QUBO problems, its scalability is limited by the number of qubits and the challenges presented by circuit depth, making it difficult to address large-scale, real-world problems on current quantum devices (Fig. 1b) \cite{preskill2018quantum, de2021materials, kim2024performance}. Note that, as seen in Fig. 1c, a quantum device (\texttt{IBM-Strasbourg}) shows potential \revnew{speedup} of QAOA jobs compared to a quantum simulator. On the other hand, a quantum emulator (\texttt{IBM-FakeBrisbane}), which simulates quantum processes considering hardware noise via a density matrix, faces significantly higher memory demands, leading to further scaling of time-to-solution compared to a quantum simulator (Fig. S1). Despite the potential advancements of quantum hardware, scalability remains limited due to the excessive number of two-qubit gates. For example, the quantum device faces difficulty in solving QUBO problems larger than 22. A `divide-and-conquer'  strategy has been proposed to address scalability challenges, where a large problem is partitioned into several sub-problems that are individually solved using QAOA \cite{dunjko2018computational, li2022large, zhou2023qaoa}. While effective for sparse problems, this approach struggles with dense (fully connected) QUBOs, which are common in real-world scenarios \cite{li2022large, zhou2023qaoa}.

We establish new policies for decomposition and aggregation procedures to efficiently tackle large and dense (fully connected) QUBO problems: 

\noindent
\textbf{(1) Decomposition policy} ensures that correlated interactions between sub-QUBOs in a original dense QUBO are adequately captured. The input QUBO with $\textit{n}$ variables is decomposed into sub-QUBOs with $\textit{k}$ variables arranged sequentially from the first to the last decision variable (Fig. S2a). This process guarantees that every decision is visited at least once by any sub-QUBO during the initialization. In the iteration process, $\textit{k}$ decision variables are randomly selected from the \textit{n} variables to form sub-QUBOs (Fig. S2b). This process enables the consideration of correlated interactions between different sub-QUBOs. As the number of iterations increases, a broader range of correlated interactions is captured, effectively preserving the critical dependencies in the original QUBO problem. \revnew{When \textit{p} sub-QUBOs are generated, each having \textit{k} variables, time complexity for the decomposition step is \textit{O(kp)}.}

\noindent
\textbf{(2) Aggregation policy} ensures that a global solution minimizes a cost function defined by the energy state of the original QUBO. The global solution is initialized during the initialization step by incrementally adding decision variables after solving sub-QUBOs. In the iteration step, each sub-QUBO, generated according to our `\textit{decomposition policy}', is solved using QAOA, and its sub-solution is used to update the cost. The cost, $\textbf{C}_{g}$, is calculated after every update, which is computationally easy and cheap. The \textit{i}$^{th}$ decision variable of the global solution $\textit{x}_{g,i}$ is replaced by the corresponding variable $\textit{s}_{l,i}$ from the sub-solution if it results in a reduced cost; i.e., $\textbf{C}_{g,s_{l,i}}$ < $\textbf{C}_{g,x_{g,i}}$, see Fig. 1d. In cases where overlapping variables between sub-QUBOs lead to conflicting decisions, the decision that minimizes the cost is retained. These iterative updates guide the global solution toward the optimal solution by continuously reducing the cost, enabling efficient convergence toward the ground truth of the original QUBO even for dense and large problems. \revnew{The aggregation step computes the global QUBO energy through bit-wise calculation, with a time complexity of \textit{O(kpn${^2}$)}.}

To accelerate this process, multi-cores/nodes of HPC systems are employed for distributed parallel computing. Each sub-QUBO is assigned to a core or node, and the global solution $\textit{x}_{g}$ is obtained by aggregating sub-solutions $\textbf{\textit{s}}$ (from \textbf{$\textit{s}_{1}$} to \textbf{$\textit{s}_{p}$}), where $\textit{p}$ is the number of sub-QUBOs. After solving all sub-QUBOs, the original large QUBO is decomposed again into $\textit{p}$ sub-QUBOs with randomly selected decision variables for subsequent iterations (Fig. S2b). Each iteration improves the global solution’s quality, converging toward a (quasi-)optimal solution (Fig. 1d). This distributed quantum-classical algorithm, which we termed Distributed QAOA (``DQAOA"), efficiently integrates quantum computing with HPC resources. It can operate on both multi-core and single-core systems. On a single core, DQAOA simplifies to ``dq-QAOA" (decomposed QUBO with QAOA), which handles one sub-QUBO per iteration ($\textit{p}=1$). However, achieving high-quality solutions in this configuration requires many iterations in this case. In contrast, DQAOA leverages distributed parallel computing to solve multiple sub-QUBOs simultaneously ($\textit{p} \gg 1$), significantly reducing the number of iterations needed to reach high-quality solutions. For initialization, DQAOA requires at least $n-k+1$ cores to ensure that each sub-QUBO can be assigned to a separate core. Since $\textit{n}$ is typically much larger than $\textit{k}$, solving an $\textit{n}$-bit problem necessitates using more than $\textit{n}$ cores for DQAOA (see the Method section for more details).

For sparse problems, the sub-QUBOs generated after partitioning a large problem may not have interactions between them (Fig. S3a). In such cases, sub-solutions can be directly aggregated to form a global solution, as overlapping variables between sub-QUBOs do not conflict due to the absence of correlated interactions (Fig. S3b). However, for denser problems, partitioning often results in sub-QUBOs with correlated interactions, complicating the simple aggregation of sub-solutions (Fig. S3c). To address this, our decomposition policy introduces additional sub-QUBOs that capture these correlated interactions (Fig. S3d). The sub-solutions from these additional sub-QUBOs are then used to correct conflicting variables, following our `\textit{aggregation policy}'. Real-world optimization problems, which feature densely (fully) connected interactions, require handling correlations across all sub-QUBOs. To handle these cases effectively, dq-QAOA and DQAOA generate numerous sub-QUBOs, determined by the number of iterations and available cores, ensuring that the global solution progressively improves in quality \revnew{through iterative bit-wise updates during aggregation}. This systematic approach effectively captures and preserves the critical dependencies of dense QUBO problems, leading to high-quality global solutions (Fig. S3e).

\subsection{Performance Analysis of DQAOA}
Given the limited access to quantum hardware, we perform the majority of our studies using the Qiskit Aer simulator, which provides near-optimal results for a single QAOA job (Fig. S1) \cite{kim2024performance}. For performance analysis, we evaluate \revnew{solution quality (1-AR value)} and time-to-solution metrics (details in Supplementary Information). While QAOA performs well on small problems using the simulator, the \revnew{solution quality decreases (i.e., 1-AR value increases)} and time-to-solution grows exponentially as problem $\textit{n}$ increases. For instance, solving even a 30-bit problem (\textit{n}=30) becomes difficult with the simulation. This exponential scaling is primarily due to the quantum simulation being executed on classical computing systems \cite{guerreschi2020intel}. In quantum simulations, all quantum states must be tracked and stored, leading to an exponential growth in memory requirements as the number of qubits required in quantum circuits increases. This challenge can be effectively mitigated by utilizing quantum hardware, where the quantum state evolution inherently leverages quantum mechanics, avoiding the exponential scaling in the time observed in classical systems (Fig. 1c). It should be noted that this work does not propose a new QAOA or simulator to improve their performances. Instead, we introduce a new strategy based on the conventional algorithm and system to scale the problem sizes while achieving a \revnew{low 1-AR value} and reduced time-to-solution. 

First, we analyze the performance of dq-QAOA for a 30-bit problem, which requires many iterations as it does not leverage distributed computing. Fig. 2a shows that increasing the number of iterations allows for achieving a \revnew{lower 1-AR value}, but it approaches convergence after 300 iterations. Notably, more iterations continue to increase time-to-solution, thus the optimal number of iterations is set to 300 for the rest of the studies unless otherwise noted. Next, we investigate the effect of sub-QUBO sizes (ranging from 4 to 10) on dq-QAOA performance. Fig. 2b indicates that while larger sub-QUBO sizes marginally improve the \revnew{solution quality}, they exponentially increase time-to-solution due to the characteristics of quantum simulation. For this study, given the use of quantum simulators, we select a sub-QUBO size of 4 for analyzing dq-QAOA.

The \revnew{performance} of dq-QAOA over standard QAOA are clearly seen in Figs. 2c and 2d. For a 30-bit problem, QAOA achieves a relatively \revnew{low solution quality with a high 1-AR of 0.16201 $\pm$0.0638}, while dq-QAOA significantly improves it to \revnew{low 1-AR of 0.05517 $\pm$0.02354}. Moreover, the time-to-solution for standard QAOA scales exponentially with the problem size, whereas dq-QAOA demonstrates a linear scaling trend \revnew{under the fixed setups (e.g., a fixed number of layers and shots for sub-QUBO solving)}, making it several orders of magnitude faster. Nevertheless, dq-QAOA yields a \revnew{high 1-AR value} for larger problems (\textit{n}$\geq$100). To address this, we tune the hyperparameters by increasing both the number of iterations and the sub-QUBO size. As shown in Figs. 2e and 2f, this hyperparameter tuning improves the \revnew{solution quality (i.e., decreases 1-AR value)} for such large problems (\revnew{0.08859 $\pm$0.04274} $\rightarrow$ \revnew{0.05299 $\pm$0.02186} for \textit{n}=100, and \revnew{0.12273 $\pm$0.02258} $\rightarrow$ \revnew{0.06082 $\pm$0.01945} for \textit{n}=150). Since decomposed sub-QUBOs collectively represent the original QUBO, using larger sub-QUBOs and a greater number of sub-QUBOs is more likely to capture essential interactions, leading to improved \revnew{solution quality}. However, the time-to-solution significantly increases ($\sim$78 \revnew{$\pm$0.45718} s $\rightarrow$ $\sim$2,134 \revnew{$\pm$29.0272} s for \textit{n}=100, and $\sim$89 \revnew{$\pm$0.33679} s $\rightarrow$ $\sim$2,405 \revnew{$\pm$12.34555} s for \textit{n}=150), which highlights the need for further improvement for the practical application to solve real-world problems.

DQAOA significantly outperforms dq-QAOA and QAOA with a \revnew{low 1-AR value (0.05453 $\pm$0.02521) and short time-to-solution (13 $\pm$3.69967 s)} for a large problem (\textit{n}=150) (Fig. 3a). Even after the hyperparameter tuning, for this large-scale problem, dq-QAOA shows a \revnew{higher 1-AR value (0.06082 $\pm$0.01945) and much longer time-to-solution (2,405$ \pm$12.34555 s)}. This notable speedup in solving large problems using DQAOA is attributed to its distributed parallel computing capabilities. By distributing sub-QUBOs across multiple cores/nodes simultaneously, DQAOA updates the global solution more frequently with a larger number of sub-QUBOs, leading to improved \revnew{solution quality}. Employing even more cores and nodes allows a greater number of sub-QUBOs (\textit{p}) for each iteration, potentially leading to a higher-quality global solution. However, Fig. 3b presents that using cores more than the problem size does not greatly improve the \revnew{solution quality}, but increases time-to-solution due to the communication overheads between cores and nodes. \revnew{This highlights a scalability consideration: beyond a certain parallelization threshold, the benefit of additional computational resources is outweighed by the cost of inter-core and inter-node communication. Our empirical analysis confirms that when the number of cores exceeds the problem size, communication delays dominate and offset the parallelism gains.} In addition, Fig. S4 reveals that increasing a sub-QUBO size (4 to 8) leads to a slightly \revnew{lower 1-AR value} but increases time-to-solution quite significantly. Hence, we set the number of iterations to 30, the sub-QUBO size to 4, and the number of cores used equal to the problem size (\textit{n}). Given these hyperparameters, DQAOA demonstrates a strong capability to scale problem sizes (\textit{n}=6 to 1,000). Notably, DQAOA achieves a \revnew{(low 1-AR value of \revnew{0.00625 $\pm$0.00122}) and short time-to-solution (276 $\pm$16.17705 s)} even for a very large problem (\textit{n}=1,000), where both QAOA and dq-QAOA cannot handle such scales (Figs. 3c and 3d). 

\subsection{Performance Comparison and Implementation on the QCSC Architecture}

While the divide-and-conquer approach with QAOA (DC-QAOA) has been proposed to address large-scale QUBO problems \cite{li2022large}, it is less effective for dense problems as its policies do not consider many correlated interactions that exist between sub-problems. In contrast, the decomposition and aggregation policies introduced in DQAOA effectively address many correlated interactions. Fig. 4a shows \revnew{relative accuracy of standard QAOA, warm-start QAOA, DC-QAOA, dq-QAOA, and DQAOA, highlighting DQAOA's superior performance, especially for large problems.} Moreover, DQAOA significantly outperforms these methods in terms of time-to-solution, with orders of magnitude faster performance (Fig. 4b). The QUBOs used in this study represent real-world problems that are both large and dense (Fig. S5), and DQAOA demonstrates the capability to solve such problems with high efficiency. \revnew{Furthermore, our recent study has shown that DQAOA can be substantially accelerated by leveraging GPU-based quantum circuit execution, further enhancing its scalability, solution quality, and real-world applicability \cite{xu2025gpu}. Classical solvers, such as parallel tempering, simulated annealing, and integer programming, can provide high-quality solutions for QUBO problems but exhibit exponentially increasing time-to-solution for large-scale problems (Fig. S6). Moreover, when addressing higher-order problems, these methods, similar to quantum annealing, often require quadratization that introduces ancillary variables, increasing computational cost and reducing overall accuracy \cite{mandal2020compressed, shaydulin2024evidence, pelofske2024short}. It is important to note that DQAOA’s flexibility enables it to efficiently solve complex higher-order binary optimization (HOBO) problems, which often exhibit rugged energy landscapes, by capturing higher-order interactions within gate-based quantum circuits while employing the same decomposition and aggregation policies \cite{pelofske2024short}. Therefore, extending DQAOA to large-scale HOBO problems represents a promising direction for future research, which will be addressed in our future work.}

DQAOA has been successfully implemented on real quantum devices integrated with HPC systems (i.e., the QCSC architecture), demonstrating the feasibility of current quantum hardware for practical applications (Fig. 4c, see details in the Method section). Hardware noise and the extensive number of two-qubit gates required for solving a QUBO with \textit{n} = 22 negatively impact algorithm performance \cite{de2021materials}. Additionally, the sparse connectivity inherent to superconducting-based quantum devices further degrades performance \cite{preskill2018quantum}, particularly for highly entangled problems common in real-world QUBO applications. Consequently, standard QAOA achieves a \revnew{high 1-AR value of 0.3259} on the \texttt{IBM-Strasbourg} quantum device. This low performance is primarily due to hardware imperfections, as evidenced by relatively \revnew{lower 1-AR value} on noiseless simulators (Fig. S1a). However, despite the constrained number of iterations posed by limited quantum device access, DQAOA and dq-QAOA achieve significantly \revnew{low 1-AR value of 0.0635 and 0.0273}, respectively, on the QCSC architecture, \revnew{under the fixed setup (e.g., a given number of layers, transpilation level, initial parameters, and classical optimizer (Fig. 4d and S7, see Methods for more details).} Notably, DQAOA yields the \revnew{lowest 1-AR} while requiring a shorter time-to-solution than dq-QAOA, emphasizing its superior performance. These results align well with those observed on emulators (\texttt{FakeBrisbane} and \texttt{FakeKyiv}) and the noiseless simulator (Fig. 4e). \revnew{We note that DQAOA can mitigate low solution quality at the algorithmic level by performing bit-wise aggregation of sub-solutions, which enhances both accuracy and robustness, thereby improving scalability even under hardware noise.}

To further evaluate scalability, we extend the problem size to a 100-bit QUBO, which is intractable for standard QAOA. Figure 4f shows that DQAOA and dq-QAOA can successfully solve this problem on the QCSC architecture. Both methods achieve \revnew{low 1-AR values} despite the restricted number of iterations, with DQAOA providing superior performance as expected (Fig. 4f). These results demonstrate DQAOA's ability to efficiently tackle large and dense real-world optimization problems that conventional QAOA cannot handle, by leveraging the QCSC architecture and employing efficient decomposition and aggregation strategies. \revnew{Moreover, the performance of DQAOA on QCSC is expected to improve significantly with next-generation quantum hardware featuring higher gate fidelity, advanced error mitigation, and ultimately error-corrected qubits. This will allow larger sub-QUBOs to be processed, reducing time-to-solution while maintaining solution quality. In addition, gate-based architectures are universal and programmable, enabling flexible circuit designs and integration with classical optimization or learning algorithms. They also can offer explicit control over circuit parameters, better scalability through distributed circuit decomposition, and a clear roadmap toward fault-tolerant quantum computation \cite{paler2017fault}. These features make gate-based approaches such as DQAOA more versatile and scalable for solving complex optimization problems.}

\subsection{Active Learning with DQAOA for Real-World Optimization Problems}
We further develop an active learning algorithm working with our DQAOA (AL-DQAOA) for optimizing material structures (layered metamaterials for spectral filters), featuring complex and large optimization spaces. This AL-DQAOA integrates machine learning, our DQAOA, and wave-optics simulation in an iterative loop (Fig. 5a) within the QCSC architecture. Here, a machine learning model (factorization machine, FM) is used to generate a QUBO as a surrogate, describing the relationship between material structures and their figure-of-merits (FOMs) \cite{kitai2020designing, kim2024quantum}. Then, DQAOA solves the given QUBO problem to predict an optimal state that represents an optimal material structure (\textbf{$\textit{x}_{po}$}). Subsequently, optical properties are evaluated by wave-optics simulation to calculate an FOM of the identified material structure (\textbf{$\textit{FOM}_{po}$}). Finally, the dataset previously used for FM training is updated by including \textbf{$\textit{x}_{po}$} and \textbf{$\textit{FOM}_{po}$}. Iterations of these processes (noted as optimization cycles) improve the dataset quality and refine QUBO surrogates, gradually converging toward the optimal state in the global optimization space. It should be noted that this algorithm is designed for minimization optimization problems, where materials with lower FOMs yield higher performances (see the Method section for more details). 

We apply AL-DQAOA to optimize spectral filters with complex optical properties depending on the wavelength of the incident light, which can be used as a transparent radiative cooler (TRC) for window applications (Figs. 5b and 5c, see the Method section for more details). AL-DQAOA can effectively find the optimal structure in a global optimization space for a 6-layered structure (12-bit optimization problem), as shown in Figs. 5d and S8. More iterations are required for larger optimization spaces, generally achieving convergence within several thousand optimization cycles (Fig. 5d). \revnew{We set the stop threshold to 90 random structure insertions among the most recent 100 optimization cycles, with a maximum of 300, 3,000, 4,000, and 5,000 optimization cycles for 12-, 30-, 50-, and 100-bit systems, respectively, reflecting the increasing size of the optimization space. The random structure insertion strategy helps avoid local optima, enabling more stable and reliable convergence toward high-quality solutions (Fig. S9 and S10).}

Here, the optimization time per cycle for 12, 30, 50, and 100-bit systems is \revnew{12.55 $\pm$0.28, 20.03 $\pm$2.59, 29.62 $\pm$0.79, and 35.28 $\pm$0.27 s}, respectively, with the simulation time using the transfer matrix method (TMM) being less than 0.1 s (Fig. 5e) \cite{luce2022tmm}. Note that DQAOA takes 7 to 10 s and the total time required for completing optimization of 12, 30, 50, and 100-bit systems are respectively \revnew{0.93 $\pm$0.14}, \revnew{12.5 $\pm$7.56}, \revnew{32.9 $\pm$1.08}, and \revnew{49.1 $\pm$0.16} hours, which should be much faster than AL with dq-QAOA (AL-dq-QAOA). To further demonstrate the superior performance of DQAOA, we optimize the same systems using dq-QAOA instead of DQAOA within AL. The results show that optimization using AL-dq-QAOA results in higher FOMs (indicating lower performances) and longer optimization times since dq-QAOA has a \revnew{higher 1-AR} and takes longer in solving given QUBOs than DQAOA (Figs. S11 and S12), which agrees well with our performance analysis study in Figs. 2 and 3. These results highlight the benefit of using DQAOA within the AL algorithm, not just for reducing optimization time but also for identifying high-performance materials. \revnew{Although conventional QUBO solvers, such as quantum annealing \cite{kim2022high, kim2024quantum} and simulated annealing (Fig. S13), can also be integrated within AL and yield reasonable optimization results, materials optimization problems involving complex, many-variable interactions (i.e., HOBOs) cannot be efficiently addressed by such solvers \cite{mandal2020compressed, pelofske2024short}. In contrast, DQAOA can provide a scalable quantum–classical framework capable of efficiently handling these higher-order interactions, as mentioned, presenting a promising direction for extending AL-DQAOA to tackle higher-order materials optimization challenges in future work.}

The optimal structure identified by AL-DQAOA has a low FOM of 0.5208 (Fig. 5d). Hence, as shown in Fig. 5f, the optimized material exhibits desired optical properties close to the ideal one (high transmitted solar irradiance in the visible range and low in the ultraviolet/near-infrared ranges). Furthermore, due to the thermal radiative layer on the top, this material yields high emission efficiency in the mid/long infrared range (Fig. S14) \cite{zhu2021subambient}. These distinct optical properties can lead to energy-saving capability for cooling by reflecting heat-generating photons and emitting thermal radiation through the atmospheric window while transmitting visible light \cite{kim2021visibly, kim2022high}. Thus, it can be used for TRC windows, which have attracted a lot of attention recently in addressing the global warming issue \cite{li2019radiative, wang2021scalable}. We calculate the energy-saving capability when using this optimized material for windows (see more details in Supplementary Information). The calculation results reveal that the optimized spectral filter used as TRC windows can save cooling energy consumption by up to $\sim$34.3\% compared to conventional windows (Figs. 5g and S15). The results demonstrate that AL-DQAOA allows for systematic design and optimization, leveraging the capabilities of both HPC and QC, contributing to the development of high-performance materials with complex, large optimization spaces.

\section{Discussion}
In this work, we have developed DQAOA to solve large-scale, real-world optimization problems based on conventional quantum algorithms and current quantum computing systems. Our performance analysis demonstrates that DQAOA achieves a high \revnew{solution quality} and significantly reduces time-to-solution by employing distributed computing across multi-cores and nodes on the HPC-QC integrated system. Consequently, DQAOA effectively handles extensive problems, such as a 1,000-bit problem, achieving \revnew{a high solution quality with an 1-AR of 0.00625 $\pm$0.00122} and time-to-solution of \revnew{276 $\pm$16.17705 s}, which conventional methods may find challenging to solve. Furthermore, we demonstrate the implementation of DQAOA on the QCSC architecture, achieving a high-quality solution. To extend its application in material science, we have proposed AL-DQAOA, an advanced optimization algorithm aiming to optimally design material structures using gate-based QC. This algorithm successfully optimizes photonic structures (spectral filters for TRC as examples), representing 100-bit real-world optimization problems. We expect that the proposed DQAOA working with the QCSC architecture can be generalized to various optimization tasks, such as finance, routing, scheduling, and assignment. In addition, the proposed AL-DQAOA will be useful for material design in general. Future enhancements could include adopting advanced quantum algorithms and next-generation quantum computing systems, including \revnew{GPU-accelerated simulators} and devices, to further optimize DQAOA's performance and expand AL-DQAOA's applicability in material design and beyond.


\section{Methods}\label{sec2}
\subsection{QAOA}
QAOA is a hybrid quantum-classical algorithm designed to solve combinatorial optimization problems by leveraging quantum systems. For minimization tasks, the objective is to prepare a quantum state that approximates the lowest energy state of a cost Hamiltonian (\(H_C\)), corresponding to the optimal solution of the classical problem. To achieve this, the classical problem is encoded into \(H_C\), where binary decision variables (\(d_i \in \{-1, 1\}\)) are mapped to the eigenvalues of Pauli-\(Z\) operators (\(\sigma^z\)). The algorithm starts with an initial quantum state (\(|\psi_0\rangle = |+\rangle^{\otimes n}\)) as a uniform superposition using Hadamard gates applied to all qubits. QAOA then alternates between applying two operators: one derived from the problem Hamiltonian (\(H_C\)) and the other from a mixer Hamiltonian (\(H_M = \sum_i \sigma^x_i\)), which is a parameterized quantum state \cite{zhou2020quantum, farhi2016quantum}:  
\begin{equation}
    {  
|\psi(\beta, \gamma)\rangle = e^{-i\beta_l H_M} e^{-i\gamma_l H_C} \cdots e^{-i\beta_1 H_M} e^{-i\gamma_1 H_C} |\psi_0\rangle ,
    } 
\label{eqs1}
\end{equation}
\noindent
where \(\beta = (\beta_1, \beta_2, \dots, \beta_l)\) and \(\gamma = (\gamma_1, \gamma_2, \dots, \gamma_l)\) are variational parameters, and \(l\) is the number of layers. These parameters are optimized classically to minimize the expected value of the cost Hamiltonian:  

\begin{equation}
    {  
f(\beta, \gamma) = \langle \psi(\beta, \gamma) | H_C | \psi(\beta, \gamma) \rangle.
    } 
\label{eqs2}
\end{equation}
\noindent
As \textit{l} increases (\(l \to \infty\)), QAOA can theoretically converge to the optimal solution. Increasing \(l\) generally improves the solution quality but also increases the complexity of parameter optimization and susceptibility to hardware errors. Hence, in practice, \(l\) is limited by the trade-off between the achieved approximation quality and the computational or hardware constraints \cite{khairy2020learning, zhou2020quantum}. 



\subsection{QUBO - Representing Optimization Problems}
In the field of quantum computing, variational quantum algorithms have shown great promise in efficiently addressing optimization problems, particularly when formulated with QUBO models. Then, the quantum systems rapidly identify an optimal binary state ($\bar{\textbf{\textit{x}}}$) that corresponds to the ground state (min $\bar{y}$) of a given QUBO, as the following equation \cite{kim2024wide, kim2024quantum}:
\begin{equation}
\bar{\textbf{\textit{x}}} = {\rm arg ~ \min_{\textit{x}}} ~ \bar{y},
\label{eqn3}
\end{equation}
\begin{equation}
 \bar{y} = \sum_{\textbf{\textit{x}}_{i} \in \{0,1\}^n} \textbf{\textit{x}}^T \mathbf{Q} \textbf{\textit{x}}.
\label{eqn4}
\end{equation}
Here, \textbf {Q} and \textbf{\textit{x}} denote a QUBO matrix and binary vector, respectively. $\bar{y}$ indicates the energy state of a QUBO with a given \textbf{\textit{x}}. Various approaches have been proposed to address QUBO problems. Among them, QAOA, working with gate-based quantum computing, has demonstrated its potential in solving these optimization problems compared to its classical counterparts \cite{shaydulin2024evidence}. 

\revnew{Hybrid quantum annealing, recognized as one of the most effective solvers for large and dense QUBO instances (though it does not guarantee the true optimum), is employed to obtain reference energy values for the represented QUBOs \cite{kim2025quantum}. }

\subsection{Building QUBOs}
To solve optimization problems for material design using quantum computing, surrogates need to be defined with QUBO formulations, which describe the relationship between material structures and their FOMs. In this study, factorization machine (FM), a supervised learning algorithm, is used as a surrogate model to build QUBOs that represent optimization problems. FM efficiently learns the relationship between input binary vectors \textbf{\textit {x}} and corresponding output values $y$ within a training dataset by the following equation \cite{kitai2020designing, kim2022high, kim2024review}: 
\begin{equation}
y = w_0 + \sum_{i=1}^{n} w_i {\textit{x}}_i + \frac{1}{2} \sum_{f=1}^m \left[
\left( \sum_{i=1}^n v_{i,f} {\textit{x}}_{i} \right)^2 - \sum_{i=1}^{n} v_{i,f}^2 {\textit{x}}_{i}^2 \right].
\label{eqn2}
\end{equation}
Here, $n$ is the length of the input binary vector \textit {\textbf {x}}, and $m$ denotes the latent space size. The length of the input vector is denoted as the problem size \textit{n} (i.e., QUBO size, $\textit{n}\times\textit{n}$ matrix). FM model parameters ($w_0, w_i$, and $v_{i,f}$) can be obtained after training, where $w_0, w_i$, and $v_{i,f}$ respectively indicate a global bias, a linear coefficient for self-interaction, and a quadratic coefficient for pairwise interactions between features. These model parameters are directly mapped to a QUBO model, which serves as a surrogate by describing the relationship between material structures and their FOMs \cite{kim2022high, kim2024review}. 

To analyze the performance of QAOA, dq-QAOA, and DQAOA, we generate QUBO problems after training FM with datasets that depict the relationship between material structures (spectral filters) and their calculated FOMs. Hence, these QUBO matrices are large and dense (Fig. S5), reflecting real-world optimization scenarios. In this study, the QUBO size (i.e., the length of a binary bit string representing a material structure) ranges from 6 to 1,000, indicating 6-bit to 1,000-bit problems.

\subsection{QUBO Decomposition}
Large QUBO problems are hard to solve with current conventional quantum technologies. Therefore, we decompose a large QUBO into smaller sub-QUBOs and aggregate sub-solutions from the sub-QUBOs to obtain a global solution. This decomposition strategy greatly reduces the complexity of handling a large QUBO. For example, a QUBO problem with 8 decision variables involves 8 self-interactions and 28 pairwise interactions. By fixing 3 decisions and decomposing the QUBO into several sub-QUBOs with 5 decisions each, we can reduce the complexity of each sub-QUBO to have 5 self-interactions and 10 interactions (Fig. 1d), making it more suitable for current quantum computing systems. This QUBO decomposition can find a true global optimal solution of an original large QUBO when preserving essential structure and interactions of the original problem \cite{glover2019quantum, atobe2021hybrid}. However, it can be computationally expensive and ineffective to solve all sub-QUBOs to identify the global optimal solution, especially for large-scale problems due to the exponential number of possible combinations ${}_\textbf{\textit{n}}C_\textbf{\textit{k}}$. Here, $\textit{n}$ and $\textit{k}$ represent the size of the original QUBO and sub-QUBOs, respectively. Hence, our algorithm iterates the decomposition and aggregate processes with randomly selected sub-QUBOs among the whole possibilities. Then, sub-QUBOs are solved using DQAOA.

Divide-and-conquer methods with QAOA (DC-QAOA) have been proposed to tackle large-scale problems by partitioning them into smaller sub-problems. However, these strategies are primarily designed for sparse problems, such as Max-Cut, with the lack of considering correlated interactions between sub-problems. \revnew{These approaches often follow a branch-and-bound-like structure, recursively partitioning and aggregating sub-problems. As a result, they are less effective for densely connected problems, where all nodes interact with one another, and important correlations between variables can be lost. In contrast, DQAOA is specifically designed for dense QUBOs. It forms multiple overlapping sub-QUBOs, each sharing some nodes and interactions with others, ensuring that essential correlations are preserved. Aggregating the solutions from these overlapping sub-QUBOs yields high-quality global solutions. This overlapping sub-QUBO formulation, together with distributed quantum-classical processing and aggregation that preserves key interactions, is key to DQAOA's superior performance on large and dense real-world problems, as shown in Figs. 4a,b and S3 \cite{dunjko2018computational, li2022large, zhou2023qaoa}.}
 
For a comprehensive comparative analysis of \revnew{solution quality} and time-to-solution across different methods, we formulate five distinct QUBOs with random elements following a Gaussian distribution (mean = 0, standard deviation = 0.15). These QUBOs exhibit similar element distributions to those obtained from the FM training with material data, effectively representing the characteristics of real-world optimization problems (Fig. 4a,b, and Fig. S5). For DC-QAOA, the sub-problem size that yields the highest \revnew{solution quality} is selected from the range of 4 to 22, leading to variations in time-to-solution based on the chosen sub-problem size.

\subsection{HPC Systems for Computational Experiments}
Our algorithm is tested on the Frontier supercomputer at the Oak Ridge Leadership Computing Facility, which is equipped with 64-core AMD ``Optimized 3rd Gen EPYC'' CPUs and 512 GB of memory per compute node \cite{budiardja2023ready}. In addition, each core interfaces with quantum systems (specifically, Qiskit Aer simulator in this study). dq-QAOA runs with a single core while DQAOA exploits multiple cores and nodes depending on the problem size. In this work, we use 50 cores per node, meaning that 150-bit problems would require 150 cores across 3 compute nodes. Distributed tasks across multi-cores and nodes are achieved by a message passing interface (MPI) implementation using mpi4py, enabling the simultaneous handling of numerous sub-QUBOs in each iteration \cite{dalcin2011parallel}. 

For the implementation of dq-QAOA and DQAOA on the QCSC architecture, we utilize the Defiant system, which provides the extended wall times and reduced queuing times. The Defiant system features a configuration similar to Frontier, equipped with 64-core AMD EPYC 7662 ``Rome'' CPUs and 256 GB of memory per compute node.

\subsection{Quantum Hardware}
To use quantum hardware, we utilize a new version of qiskit (1.2.4) and qiskit-ibm-runtime (0.32.0) to interface with \texttt{IBM quantum devices} as backends. \revnew{The original simulator-based experiments are performed using Qiskit 0.41.0; however, hardware experiments on the QCSC require Qiskit 1.2.4 and qiskit-ibm-runtime 0.32.0, as earlier versions do not currently support quantum circuit execution on real devices.} The QAOA ansatz is constructed using the \textit{QAOAAnsatz} function and subsequently transpiled using a \textit{generate-preset-pass-manager} provided by qiskit. \revnew{The number of QAOA layer is set to 1, the optimization level for circuit transpilation is set to 2, initial parameters are set to $\pi/4$ for $\gamma$ and $\pi/8$ for $\beta$, and COBLYA (a gradient-free optimizer) is employed as a classical optimizer, yielding high solution quality with a reduced time-to-solution under these conditions (Fig. S7).} 

We use \texttt{FakeBrisbane} and \texttt{FakeKyiv} as emulated backends (fake backends), which simulate the behavior of a quantum device while incorporating the effects of hardware errors due to noise. For quantum hardware, we utilize the devices \texttt{IBM-Strasbourg} and \texttt{IBM-Kyiv}, each featuring 127 qubits. Among these devices, the least busy device is selected for quantum circuit execution. The time-to-solution is calculated as the sum of quantum processing unit (QPU) usage time and parameter optimization time on classical hardware, explicitly excluding queueing delays. The time-to-solution on the simulators and emulators scales exponentially with problem size. In contrast, the time-to-solution of quantum hardware exhibits linear scaling, \revnew{under the fixed setups (e.g., fixed number of layers, transpilation level, number of shots, and backends)} (Fig. 1c). \revnew{Currently, FakeStrasbourg is not available in the ibm-runtime-service library. Therefore, we used FakeBrisbane, which is based on the Eagle r3 processor, the same generation as IBM Strasbourg, ensuring comparable noise characteristics and hardware topology.}

We demonstrate the performance of dq-QAOA and DQAOA on \texttt{IBM-Strasbourg} and \texttt{IBM-Kyiv}, which provide a direct interface between HPC cores/nodes and quantum computing systems, enabling the distribution of sub-QAOA tasks across a quantum-centric supercomputing architecture (Fig. 4c). Hardware noise, particularly affecting two-qubit gates, makes solving QUBOs with \textit{n} $>$ 22 on IBM quantum devices challenging, often resulting in poor performance \revnew{characterized by a high 1–AR value}. Emulators are similarly constrained, being difficult to handle QUBOs with \textit{n} $>$ 12 due to exponential memory demands inherent in density matrix simulation. To address these constraints, we solve a QUBO with \textit{n} = 22 with a sub-QUBO size of 11 as a proof-of-concept. Given the limited quantum hardware access, the number of iterations is restricted to 5. Directly solving the QUBO with QAOA on the quantum device results in a \revnew{high 1-AR value of 0.3260} due to hardware noise and deep circuit depth. However, the use of DQAOA and dq-QAOA significantly improves the \revnew{solution quality with high 1-AR values of 0.0273 and 0.0635, respectively} (Fig. 4d,e). These findings highlight the practical feasibility of our distributed approach for addressing large and dense optimization problems on quantum hardware (Fig. 4d). For large-scale testing (\textit{n} = 100), we omit the initialization step in both dq-QAOA and DQAOA to manage quantum computing costs. Instead, the global solution is initialized with random binary variables (Fig. 4f), and we extend the number of iterations up to 40. This approach can result in high \revnew{solution quality}.

\subsection{AL-DQAOA}
Our AL-DQAOA consists of three key components: 

\noindent  \textbf {(1) Factorization Machine:} FM is trained with a dataset to construct a QUBO model that serves as a surrogate by describing the correlation between material structures and their FOMs. 

\noindent  \textbf {(2) DQAOA:} DQAOA is employed to identify the approximated optimal binary state corresponding to the ground state of a QUBO model, which is formulated with the FM model parameters.

\noindent  \textbf {(3) Wave-Optics Simulation:} The optical properties of the approximated optimal material state obtained from DQAOA (in step 2) are evaluated to calculate a FOM. Here, we use TMM for evaluating the optical properties of the designed materials, but other types of simulation tools can be used for different purposes. 

The dataset is updated by adding the identified optimal binary state (from step 2) and the calculated FOM (from step 3). Iterating these processes improves the dataset quality, enabling FM to build a more reliable surrogate. Subsequently, DQAOA can discover a higher-quality data point, which leads to the convergence toward the optimal state in a global optimization space as iterations progress (see Fig. S10). DQAOA can select the same binary vector that was already identified in a previous optimization cycle. In this scenario, a randomly generated binary vector (i.e., random material structure) and its calculated FOM are added for updating the dataset instead of those obtained from DQAOA to avoid redundancy and increase diversity in the dataset (red dots in Figs. S10 and S12).

We first generate an initial dataset that includes 25 random photonic structures and their calculated FOMs, and start optimization with AL-DQAOA. Before training FM, we split a dataset into a training set and a test set in a 4:1 ratio. Random structures are increasingly added to the dataset, meaning that DQAOA identifies the repeated structures, when the convergence is almost reached. \revnew{Without this random structure insertion strategy, which prevents repeated structures from the dataset, the active learning process can easily become trapped in local optima, hindering convergence to high-quality optimization solutions \cite{kim2022high}. As shown in Fig. S9, removing random structure insertion leads to lower solution quality, evidenced by higher minimum FOMs.} In addition, the convergence is usually achieved within several thousand optimization cycles, depending on the problem sizes. Upon reaching convergence, AL-DQAOA usually does not find higher-quality data points \cite{kim2022high, kim2024wide}. Therefore, we stop iterations upon satisfying preset thresholds to save computational costs: (1) 90 data out of the last 100 optimization cycles are from random structures, which indicates that the convergence is nearly achieved. (2) The number of optimization cycles reaches \revnew{300,} 3,000, 4,000, and 5,000 for \revnew{12-bit,} 30-bit, 50-bit, and 100-bit problems, respectively\revnew{, considering exponentially enlarging optimization spaces with the system size.} Given these strategies and thresholds, we can successfully optimize material structures (in this study, spectral filters as examples), featuring significantly large optimization spaces up to $2^{100}$.

\subsection{Optimal Design of Spectral Filters using AL-DQAOA}
We apply our AL-DQAOA to optimize spectral filters, and our goal is to optimally design spectral filters for high transmission in the visible spectrum while attenuating transmission in the ultraviolet (UV) and near-infrared (NIR) regions. Additionally, a thermal radiative polymer layer is added on the top of the filter surface, focusing on their application in the transparent radiative cooler (TRC). \cite{kim2022high}. TRC has attracted a lot of interest recently due to its potential to address the global warming issue by reducing cooling energy consumption \cite{wang2021scalable, kim2021visibly, kim2022high}. This energy-saving capability comes from the distinct optical properties of the designed spectral filter to reflect heat-generating photons (UV and NIR photons) while selectively transmitting visible light. Additionally, the polymer layer on the top emits thermal radiation through the atmospheric window \cite{zhu2021subambient}. 

We design planar multilayered structures (PML) for spectral filters (Fig. 5b). The PML consists of 3 to 50 layers (total thickness is 1,200 nm), and each layer is composed of one of four material candidates: silicon dioxide: SiO$_2$, silicon nitride: Si$_3$N$_4$, aluminum oxide: Al$_2$O$_3$, and titanium dioxide: TiO$_2$. With each layer represented by a two-digit binary label (‘00’ for SiO$_2$, ‘01’ for Si$_3$N$_4$, ‘10’ for Al$_2$O$_3$, and ‘11’ for TiO$_2$), a layer configuration can be described by a binary vector. Here, optimizing \textit{nl}-layered spectral filters represents 2\textit{nl}-bit optimization problems, thus optimizing 50-layered spectral filters represents a 100-bit optimization problem. For example, a 6-layered structure composed of SiO$_2$/Si$_3$N$_4$/Al$_2$O$_3$/TiO$_2$/Al$_2$O$_3$/Si$_3$N$_4$ is represented as [00 01 10 11 10 01]. Spectral filters for TRC can be realized by adding 40 $\mu$m of polydimethylsiloxane layer on the top of the PML. The ideal spectral filter (i.e., ideal TRC) should have unity transmission in the visible regime and zero transmission in the UV/NIR regimes, resulting in a high solar-weighted transmission (i.e., transmitted solar irradiance) in the visible range and zero in other ranges. The objective is to align the optical properties of the designed spectral filters as closely as possible with these ideal properties in terms of transmitted irradiance. To evaluate the performance of the designed spectral filters, we calculate the FOM using the following equation:

\begin{equation}
\text{FOM} = \frac{ 10\int_{\lambda=300}^{\lambda=2,500} [(T_{ideal}(\lambda)S(\lambda))^2 - (T_{designed}(\lambda)S(\lambda))^2]d\lambda} {\int_{\lambda=300}^{\lambda=2,500} S(\lambda)^2 d\lambda} .
\label{eqn1}
\end{equation}

In this equation, $S(\lambda)$ is the solar irradiance, $T_{designed}(\lambda)$ and $T_{ideal}(\lambda)$ indicate the transmission efficiency of a designed and ideal spectral filter, respectively. $T(\lambda)S(\lambda)$ denotes the transmitted irradiance. The optical properties ($T_{designed}(\lambda)$) are calculated using the transfer matrix method, a highly effective optical simulation tool for multilayered structures \cite{kim2022high, kim2024wide, luce2022tmm}. It is important to note that this represents a minimization optimization problem, as the FOM approaches 0 when the spectral filter exhibits high performance having optical properties similar to the ideal one.


\makeatletter
\setcounter{figure}{0}

\makeatother

\section*{Acknowledgments}
The authors thank the Materials Science Working Group for useful discussions. This research used resources of the Oak Ridge Leadership Computing Facility at the Oak Ridge National Laboratory, which is supported by the Office of Science of the U.S. Department of Energy under Contract No. DE-AC05-00OR22725. This material is based upon work supported by the U.S. Department of Energy, Office of Science, National Quantum Information Science Research Centers, Quantum Science Center.
{\it Notice}: This manuscript has in part been authored by UT-Battelle, LLC under Contract No. DE-AC05-00OR22725 with the U.S. Department of Energy. The United States Government retains and the publisher, by accepting the article for publication, acknowledges that the U.S. Government retains a non-exclusive, paid up, irrevocable, world-wide license to publish or reproduce the published form of the manuscript, or allow others to do so, for U.S. Government 15 purposes. The Department of Energy will provide public access to these results of federally sponsored research in accordance with the DOE Public Access Plan (http://energy.gov/downloads/doe-publicaccess-plan). 

\section*{Data availability}
The data used for this work can be found at https://github.com/Seongmin-Kim-ornl/DQAOA. 

\section*{Code availability}
The codes developed for this work are available at https://github.com/Seongmin-Kim-ornl/DQAOA.

\section*{Author information}
Authors and Affiliations:

\noindent
{\bf National Center for Computational Sciences, Oak Ridge National Laboratory, Oak Ridge, Tennessee 37830, United States.}

Seongmin Kim, Zhihao Xu \& In-Saeng Suh

\noindent
{\bf IBM Quantum, IBM T.J. Watson Research Center, Yorktown Heights, New York 10598, United States.}

Vincent R. Pascuzzi

\noindent
{\bf  Department of Aerospace and Mechanical Engineering, University of Notre Dame, Notre Dame, Indiana 46556, United States}

Zhihao Xu \& Tengfei Luo

\noindent
{\bf Department of Electronic Engineering, Kyung Hee University, Yongin-Si, Gyeonggi-do 17104, Republic of Korea}

Eungkyu Lee

\section*{Contributions}
S.K. and I.S. conceived the idea and designed the experiments. S.K. developed the codes, derived all experimental results, and performed data analyses. All authors discussed the results, and contributed to the writing of the manuscript.

\section*{Corresponding authors}
Correspondence to In-Saeng Suh, Eungkyu Lee, Tengfei Luo, and Seongmin Kim.

\noindent 
Email address: In-Saeng Suh (suhi@ornl.gov), Eungkyu Lee (eleest@khu.ac.kr), Tengfei Luo (tluo@nd.edu), and Seongmin Kim (kims@ornl.gov).

\section*{Ethics declarations}
\textbf{Competing Interests}\\
The authors declare no competing interests.

\printbibliography
\clearpage
\section*{Figures}
\begin{figure}[!ht]
\centering
\includegraphics[width=1.0\linewidth]{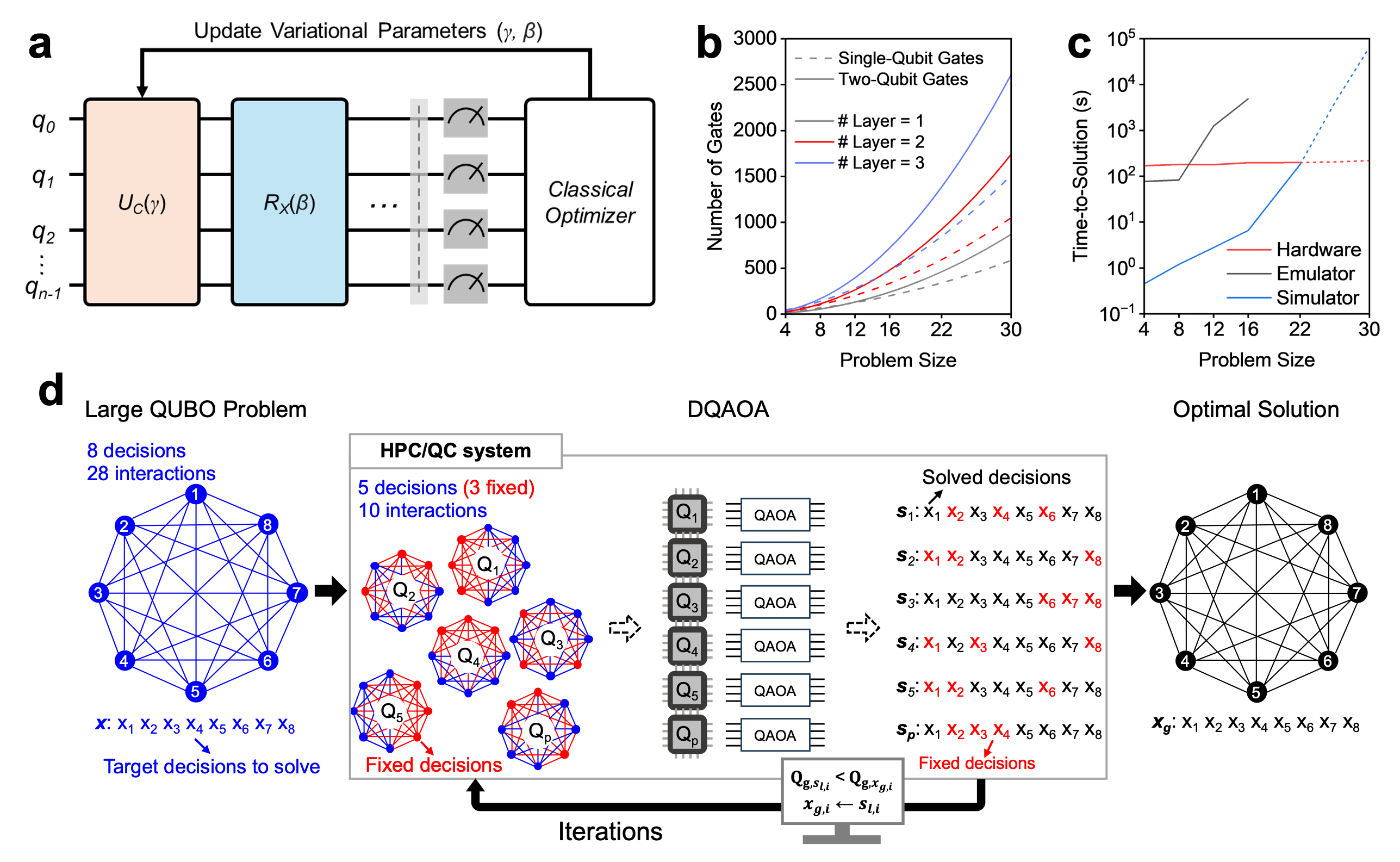}
\caption{\label{fig:fig-1} \textbf{Workflow of DQAOA leveraging a quantum-centric supercomputing architecture.} \textbf{a}, The schematic of a QAOA circuit, illustrating the iterative process to update variational parameters in both classical and quantum components. \textbf{b}, The number of single-qubit and two-qubit gates as a function of QUBO problem size, indicating that circuit depth grows significantly for larger problems, reflecting the increased complexity of solving large-scale optimization tasks. \textbf{c}, Time-to-solution of QAOA for QUBO problems on quantum hardware (\texttt{IBM-Strasbourg}), an emulator (\texttt{IBM-FakeBrisbane}), and a simulator (\texttt{Qiskit-Aer}). The dotted lines indicate the expected time-to-solution for solving large QUBOs (\textit{n} $>$ 22). \textbf{d}, The schematic of DQAOA to solve large-scale optimization problems through distributed computing. A large QUBO is decomposed into \textit{p} sub-QUBOs, which are solved by the quantum-centric supercomputing architecture.}
\end{figure}

\clearpage
\begin{figure}[!ht]
\centering
\includegraphics[width=1.0\linewidth]{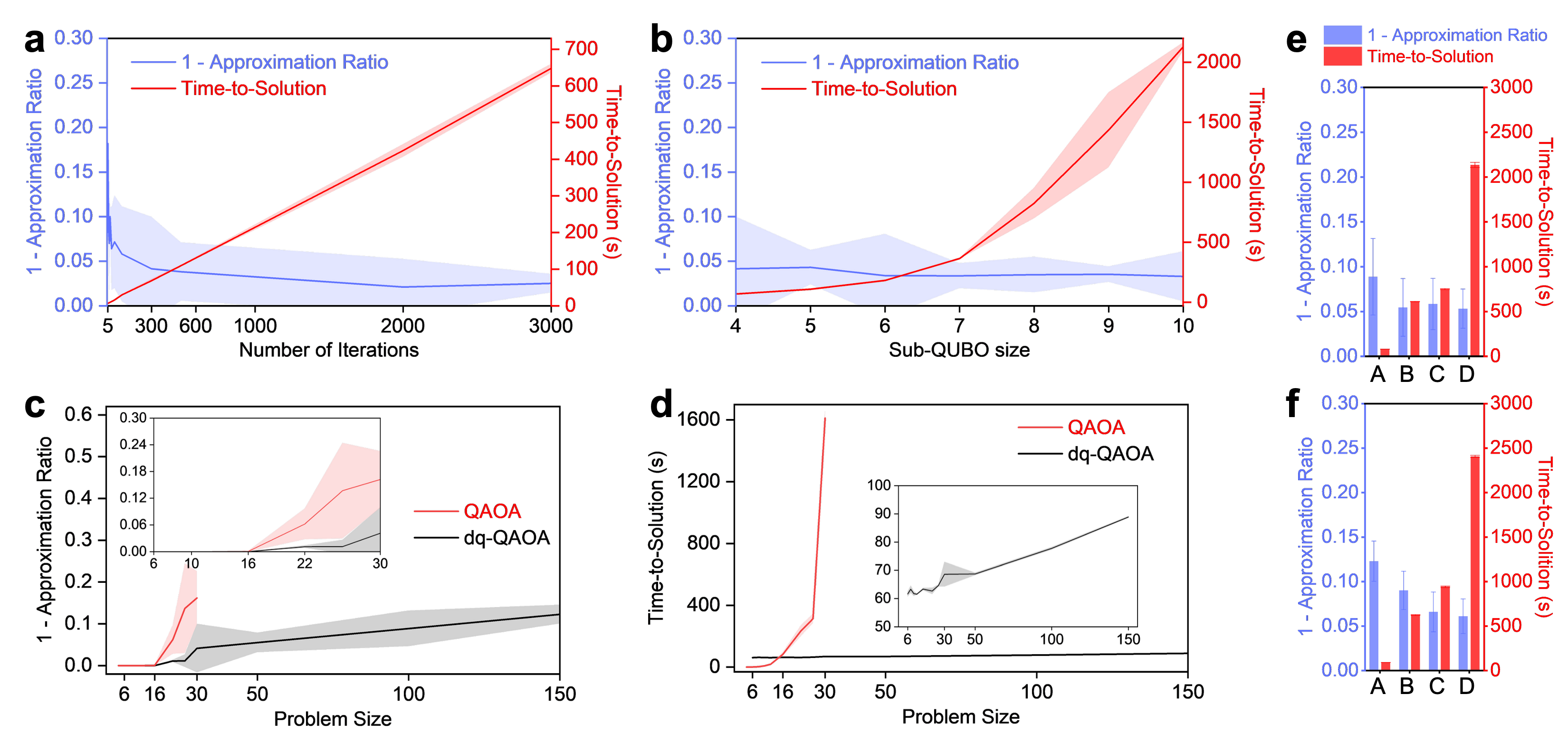}
\caption{\label{fig:fig-2} \textbf{Performance analysis of dq-QAOA.} \revnew{1-AR} and time-to-solution for a 30-bit problem as a function of (\textbf{a}) the number of iterations, and (\textbf{b}) the sub-QUBO size. The number of iterations is set to 300 and sub-QUBO size to 4 for further studies, as \revnew{1-AR value} nearly converges but time-to-solution continues to increase. (\textbf{c}) \revnew{1-AR} and (\textbf{d}) time-to-solution of QAOA and dq-QAOA as a function of the problem size. \revnew{1-AR} and time-to-solution of dq-QAOA for a problem size of (\textbf{e}) 100 and (\textbf{f}) 150 with the different number of iterations and sub-QUBO sizes. \textbf{A}: 300 iterations with the sub-QUBO size of 4. \textbf{B}: 3,000 iterations with the sub-QUBO size of 4. \textbf{C}: 300 iterations with the sub-QUBO size of 8. \textbf{D}: 1,000 iterations with the sub-QUBO size of 8.}
\end{figure}

\clearpage

\begin{figure}[!ht]
\centering
\includegraphics[width=1.0\linewidth]{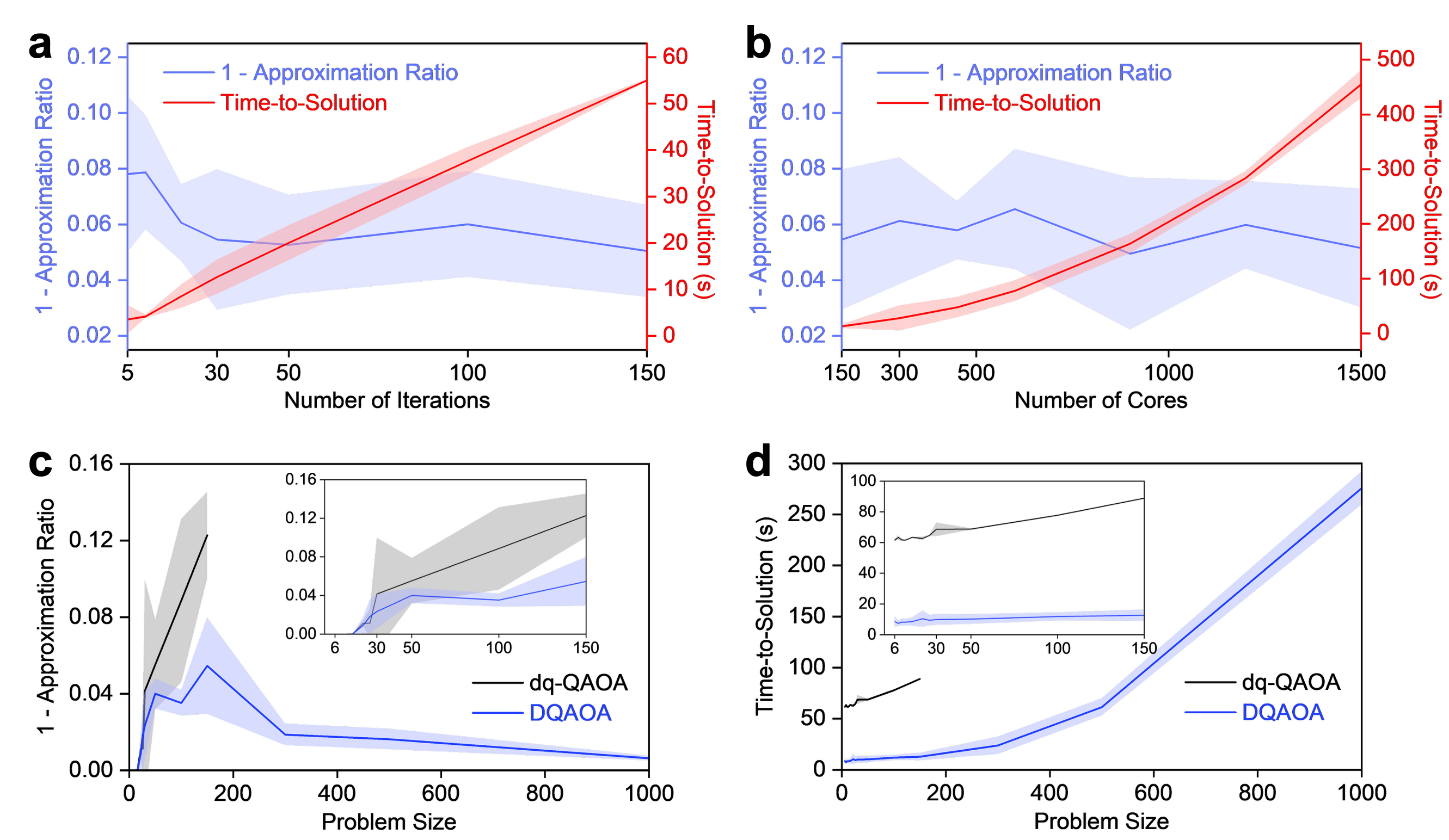}
\caption{\label{fig:fig-3} \textbf{Performance analysis of DQAOA.} \revnew{1-AR} and time-to-solution for a 150-bit problem as a function of (\textbf{a}) the number of iterations, and (\textbf{b}) the number of cores used. (\textbf{c}) \revnew{1-AR} and (\textbf{d}) time-to-solution of dq-QAOA and DQAOA as a function of the problem size (\textit{n}). Note that the number of iterations, sub-QUBO size, and the number of cores used for DQAOA are set to 30, 4, and \textit{n}, respectively.}
\end{figure}

\clearpage
\begin{figure}[!ht]
\centering
\includegraphics[width=1.0\linewidth]{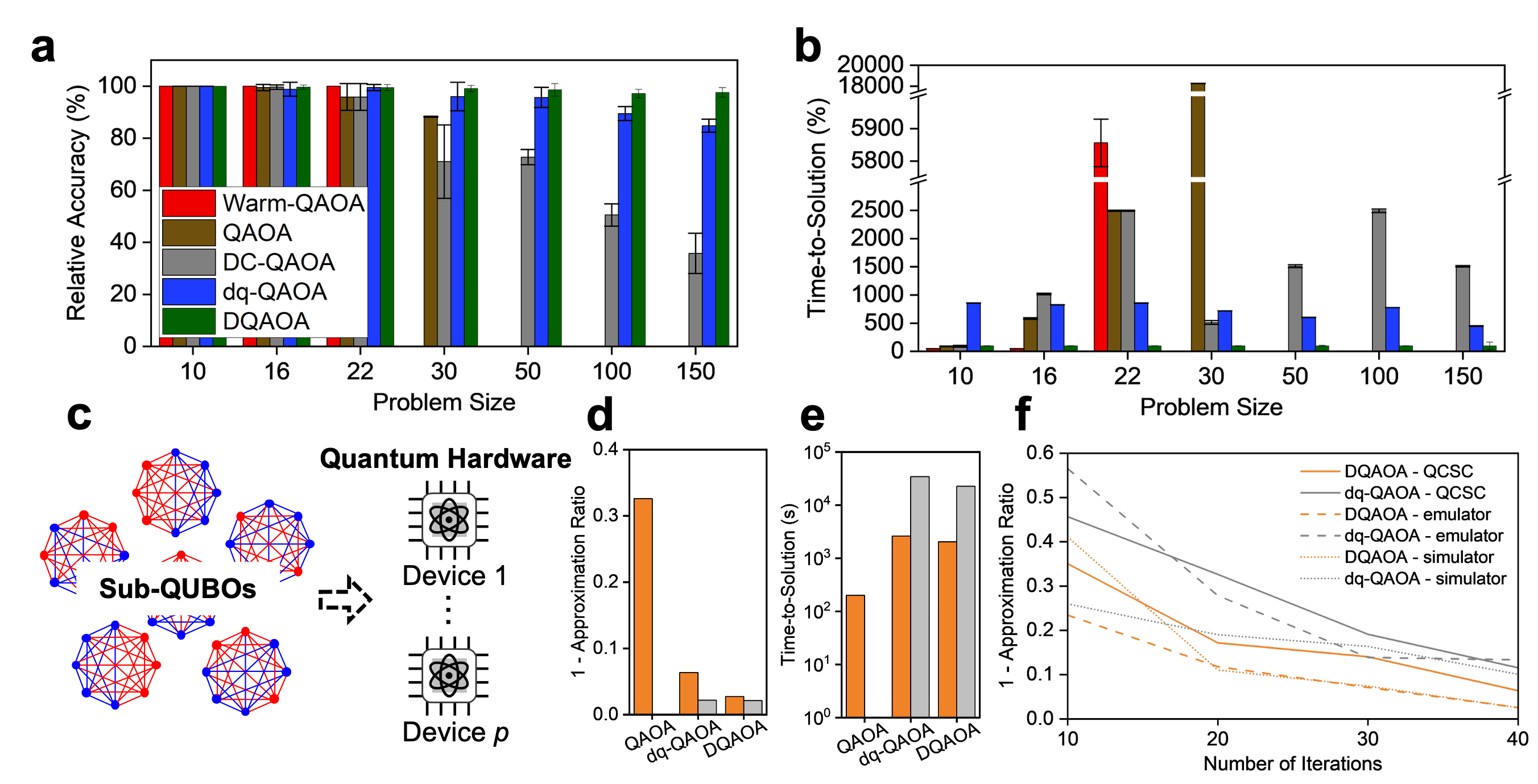}
\caption{\label{fig:fig-4} \textbf{Performance analysis of dq-QAOA and DQAOA, and hardware implementation.} (\textbf{a}) \revnew{relative accuracy} and (\textbf{b}) time-to-solution of warm-start QAOA, standard QAOA, DC-QAOA, dq-QAOA, and DQAOA, with DQAOA serving as the baseline for time-to-solution. 
\textbf{c}, Schematic representation of the implementation of dq-QAOA and DQAOA on the QCSC architecture. 
For a QUBO with \textit{n} = 22, (\textbf{d}) \revnew{1-AR} and (\textbf{e}) time-to-solution of QAOA, dq-QAOA, and DQAOA on quantum devices (orange bar; \texttt{IBM-Strasbourg} and \texttt{IBM-Kyiv}) and emulators (grey bar; \texttt{FakeBrisbane} and \texttt{FakeKyiv}). 
\textbf{f}, For a QUBO with \textit{n} = 100, \revnew{1-AR} of \revnew{DQAOA (orange) and dq-QAOA (grey) on quantum devices (solid lines; \texttt{IBM-Strasbourg} and \texttt{IBM-Kyiv}), emulators (dashed lines; \texttt{FakeBrisbane} and \texttt{FakeKyiv}), and simulator (dotted lines; \texttt{Qiskit-Aer}). }
}
\end{figure}

\clearpage
\begin{figure}[!ht]
\centering
\includegraphics[width=1.0\linewidth]{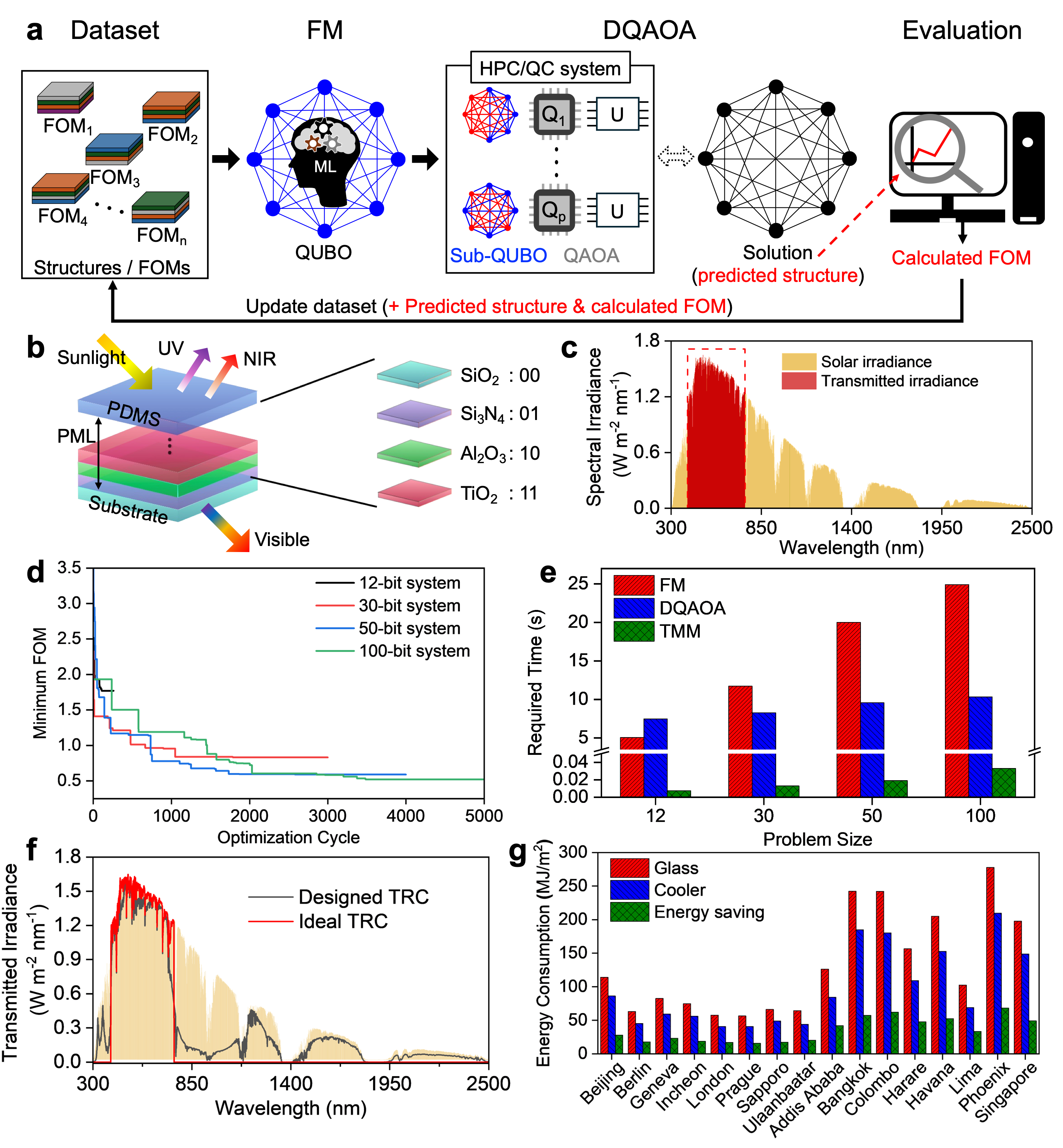}
\caption{\label{fig:fig-5} \textbf{The AL-DQAOA algorithm for solving large-scale optimization problems in material science, exemplified by spectral filters with 12, 30, 50, and 100-bit systems.} \textbf{a}, The schematic of AL-DQAOA integrating machine learning, DQAOA, and performance evaluation in an iterative loop. \textbf{b}, The schematic of a spectral filter for TRC windows, where the planar multilayered structure (PML) is subject to optimization. \textbf{c}, Solar spectral irradiance (yellow shade, air mass 1.5 global) and transmitted irradiance through the ideal spectral filter (red shade). The ideal spectral filter transmits the solar spectrum only in the visible range. \textbf{d}, The evolution of FOM as a function of optimization cycles for different problem sizes. \textbf{e}, Time required for an optimization cycle of AL-DQAOA for different problem sizes. \textbf{f}, Calculated optical properties of the spectral filter optimized by AL-DQAOA. The yellow shade presents solar spectral irradiance. \textbf{g}, Energy-saving potential when using the optimized spectral filter as TRC windows in cities around the world, compared to a conventional window.}
\end{figure}

\includepdf[pages=-]{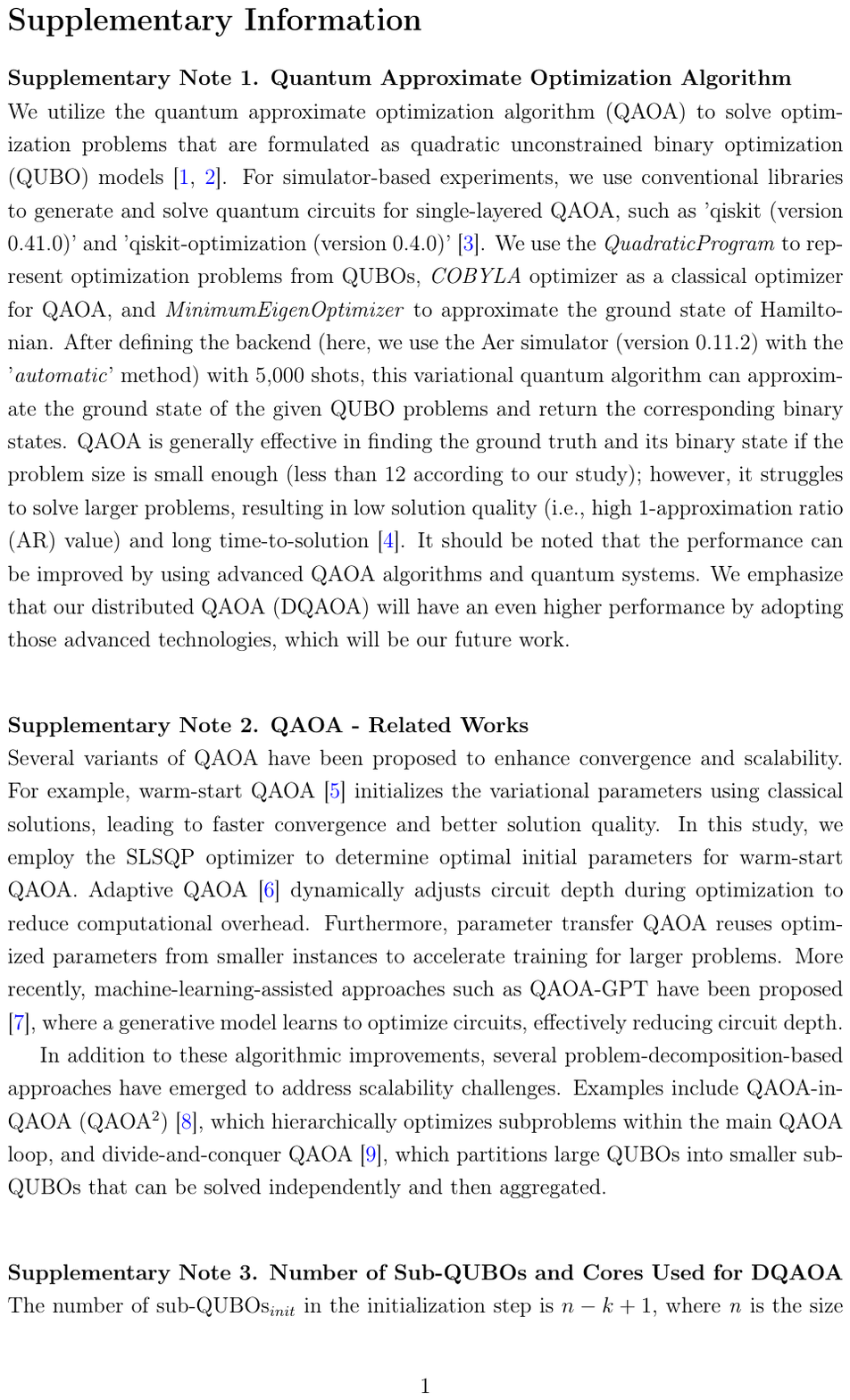}

\end{document}